# A Route to Pure Optical Rotation in Self-Assembled Materials through Energetic Non-Degeneracy

Daniel J. Gracias[1], Thomas J. Ugras[2, 3], Richard D. Robinson[1, 3, *]

[1] Department of Materials Science and Engineering, Cornell University, Ithaca, NY, USA.
[2] School of Applied and Engineering Physics, Cornell University, Ithaca, NY, USA.
[3] Kavli Institute at Cornell for Nanoscale Science, Ithaca, NY, USA.
[*] Corresponding author: rdr82@cornell.edu

## Abstract

Achieving large optical rotation with minimal ellipticity and absorption, 'pure' optical rotation, remains a central challenge in chiral photonics. Solution-processed self-assembled materials can exhibit exceptional chiroptical responses (g-factors > 1), yet their circular birefringence (CB) typically overlaps with circular dichroism (CD) and resonant loss (Absorption), leading to elliptical, attenuated signals. Here, we establish a general, theory-guided design principle showing that non-degeneracy provides a route towards pure optical rotation in self-assembled systems. Using a generalized coupled-oscillator framework, we demonstrate that breaking degeneracy between the excited states of interacting chromophores produces CB in spectral regions where CD and absorption are naturally weak. We experimentally validate this mechanism using mixed assemblies of α- and β-CdS magic-sized clusters, which exhibit the predicted off-resonant, emergent CB. Guided by this principle, we design a layered architecture that maximizes non-degenerate neighbors through alternating chromophore planes. This structural architecture results in optical response lineshapes optimized for pure rotation. Because the mechanism relies solely on dipolar coupling and energetic detuning, it is generalizable across wavelengths, including in the ultraviolet (~310 nm), where suitable nanocrystal and organic chromophores are readily available. Simulations predict a 50 meV (12 THz) window exhibiting low-dispersion optical rotation of ~20°, >40% transmission, and <1° ellipticity—strong performing benchmarks typically associated with lithographic metamaterials. These results establish non-degenerate coupling as a general mechanism for engineering chiroptical response and provide a strategy for realizing pure optical rotation in self-assembled systems.

## Main text

Achieving pure optical rotation—rotation of linearly polarized light that preserves its linearity—is essential for applications such as chiral sensing, polarization control, and integrated photonic devices [1,2]. Optical rotation of linearly polarized light can be achieved when linear light is passed through a medium that possesses circular birefringence (CB), which is a difference in refractive index between left- and right-circularly polarized light. This optical rotation, however, comes at a price: in typical materials that exhibit strong rotation, in the wavelength band of large CB, there is an accompanying strong circular dichroism (CD) resonance, where one handedness of circularly polarized light is preferentially absorbed over the other[2,3]. This proximity to a resonance leads to loss of signal, and the dichroism converts linear polarization into elliptical, limiting functionality in applications that require purely linear polarized light. Consequently, practical low-loss and low-ellipticity optical rotators must operate far from resonance, such as quartz, where weak off-resonant dispersion produces negligible absorption and CD but requires millimeter-scale thicknesses to achieve meaningful polarization rotation[4,5]. Historically, achieving pure optical rotation (rotation of linear polarization without introducing ellipticity) has been very challenging in miniaturized systems. At the micron scale, a regime desirable for integrated and on-chip components, appreciable pure rotation is only achievable in metamaterials [1], which use lithographically fabricated structures with carefully engineered resonances.

The underlying strategy used to achieve pure rotation in broadband, low-ellipticity metamaterials is remarkably consistent across architectures, independent of whether they are constructed from metals, dielectrics, or polymers[3,6–13]. These systems employ two resonances of opposite handedness (i.e., Left-hand circularly polarized and right-hand circularly polarized), separated in energy. The spectral sum of these two peaks is a lineshape with a positive and negative CD peak. As CD is related to CB through the Kramers-Kronig (KK) relations [14], the resulting CB, and thus the optical rotation, forms a broad, relatively flat response between the two resonances. Crucially, this rotation occurs in the spectral gap between the resonances, where absorption is lower, and ellipticity is minimized. This mechanism enables such chiral metamaterials to deliver appreciable pure rotation. The limitation of these types of materials, however, is that they are only realized through complex geometry and micro/nanofabrication: though they are versatile, lithographic metamaterials are fabrication-intensive and not scalable, limiting their implementation in commercial applications. In addition, their operation in the visible and ultraviolet is constrained by strong material losses and the need for extremely fine structural features at short wavelengths. These challenges highlight the appeal of passive, solution-processed materials as a scalable route to chiral photonic functionality.

In contrast to metamaterials, solution-processed self-assembled chiral materials (**Fig. 1a**) offer an appealing alternative for µm-scale optical rotation due to their economic advantages and three-dimensional scalability, but, notably, pure optical rotation has yet to be demonstrated in these systems. Large chiroptical g-factors (CD normalized by absorbance) have been reported in solution-processed thin film assemblies of organic dyes[15], semiconductor nanocrystals [16], and metal nanoparticles[17,18], across the electromagnetic spectrum (**Fig. 1b**). In these systems, intermolecular or interparticle coupling splits the chromophore transition into shifted resonances (**Fig. 1c, left**) that interact differentially with left- and right-circularly polarized light, producing the characteristic positive and negative peak (bisignate) CD spectrum[19]. The large chiroptical responses with appreciable g-factors arise from extended delocalization of collective chiral states, which is only possible in highly ordered, three-dimensionally extended structures[20–22]. The bisignate CD lineshape results, from KK relations, in a monosignate CB signal, with its maximum aligned at the zero crossing of the CD spectrum. In typical systems, however, the energy splitting between the two CD peaks is small, resulting in significant overlap between the positive and negative CD peaks. As a result, the CB peak is narrow and falls into the same spectral region as the material's resonant absorption[15,16]. The CB peak's alignment with the absorption leads to strong attenuation, large ellipticity, and consequently low-quality optical rotation (**Fig. 1d, left**).

One idea to solve this issue and achieve CD and CB profiles similar to those used for pure rotation in metamaterials is to substantially increase the energy separation between the two chiral resonances. A promising route could be to harness coupling between non-degenerate chromophores, where the inherent detuning of their transition energies can artificially enlarge the splitting between the chiral modes (**Fig. 1c, right**) [19]. While such detuning-enhanced splitting has been demonstrated in simple dimers[19,23], it has not been demonstrated in either theoretical or experimental studies for large, delocalized assemblies, where long-range coupling is essential for generating strong chiroptical responses. To date, the potential of non-degenerate coupling to engineer circular birefringence for pure optical rotation remains unexplored.

Here we show that deliberately engineered non-degeneracy provides a viable and general route to realizing pure optical rotation in self-assembled systems (**Fig. 1d, right**). Using an established coupled-oscillator formalism, we model the chiroptical response of extended assemblies composed of two non-degenerate chromophores and establish design rules for achieving a separation of circular dichroic resonances that yields a broad, off-resonant circular birefringence window. We validate this mechanism experimentally using mixed assemblies of two isomers of magic-sized clusters with ~150 meV detuning, demonstrating that non-degenerate coupling creates off-resonant CB features, desirable for pure optical rotation. Building on this understanding, we use our model to show that a non-degenerate ABAB stacking further enhances pure rotation performance. By exploring the relevant parameter space, we identify design rules that can simultaneously provide a ~20° rotation over a 50 meV bandwidth window, with <1° ellipticity and >40% transmission in an idealized structure. Importantly, this strategy is general and applies to any chromophore building block with a well-defined dipole-allowed transition, enabling the development of pure-rotation materials across a broad range of wavelengths.

**Results and Discussion**

Model Overview

A generalized coupled-oscillator formalism[19,22] is used to simulate the chiroptical response of chiral assemblies of $N$ coupled chromophores. This framework has been used in the past to model coupled chiral assemblies of organic chromophores[24], plasmonic nanorods [25], as well as semiconductor nanoclusters [16]. Every chromophore is approximated as a point dipole oscillator located at position $\boldsymbol{r}_i$, possessing a single excited state, characterized by energy $E_i$, wavefunction $\psi_i$, and a corresponding electric transition dipole moment $\boldsymbol{\mu}_i$. Coupling between two chromophores ($i$ and $j$) is assumed to be purely electromagnetic and modeled using the near-field dipole interaction potential $V_{ij}$

$$V_{ij} = \frac{1}{4\pi\varepsilon}\left(\frac{\boldsymbol{\mu}_i \cdot \boldsymbol{\mu}_j}{|\boldsymbol{r}_{ij}|^3} - \frac{3(\boldsymbol{\mu}_i \cdot \boldsymbol{r}_{ij})(\boldsymbol{\mu}_j \cdot \boldsymbol{r}_{ij})}{|\boldsymbol{r}_{ij}|^5}\right) \quad \text{(Eqn. 1)}$$

Where $\boldsymbol{r}_{ij} = \boldsymbol{r}_j - \boldsymbol{r}_i$ and $\varepsilon$ is the permittivity of the host medium. Coupling between chromophores results in the formation of $N$ hybrid delocalized states $\Psi_p$, which are eigenstates of the full interaction Hamiltonian $\widehat{H}$, with corresponding energy eigenvalues $\mathrm{E}_p$. This Hamiltonian can be written in the basis of singly excited states as an $N \times N$ matrix with diagonal elements $\hat{H}_{ii} = E_i$, and off-diagonal elements $\hat{H}_{ij} = V_{ij}$:

$$\hat{H} = \begin{bmatrix} E_1 & V_{12} & \cdots & V_{1N} \\ V_{21} & E_2 & & \\ \vdots & & \ddots & \vdots \\ V_{N1} & & \cdots & E_N \end{bmatrix} \quad \text{(Eqn. 2a)}$$

In this basis, the resulting eigenstates $\Psi_p$ are linear superpositions of the singly excited states

$$\Psi_p = \sum_i C_p^{(i)} \psi_i \quad \text{(Eqn. 2b)}$$

The energy eigenvalues $E_p$ and superposition coefficients $C_p^{(i)}$ are obtained by solving the time-independent Schrodinger equation:

$$\hat{H}\Psi_p = E_p \Psi_p \quad \text{(Eqn. 2c)}$$

For degenerate assemblies, all individual chromophores have identical excitation energies, $E_1 = E_2 = \cdots = E_N$. In contrast, for non-degenerate assemblies composed of two different chromophores, $A$ and $B$, with excited state energies $E_A$ and $E_B$. The exact ratio and positioning of $E_A$ and $E_B$ on the diagonal of the Hamiltonian depends on the composition and the arrangement of the chromophore assembly. The orientationally-averaged absorption and circular dichroism magnitudes of each of the hybrid states $\Psi_p$ are then calculated using the dipole strength ($D_p$) and rotational strength ($R_p$), given by

$$D_p = \sum_i {C_p^{(i)}}^2 |\boldsymbol{\mu}_i|^2 + 2 \sum_{i<j} C_p^{(i)} C_p^{(j)} \boldsymbol{\mu}_i \cdot \boldsymbol{\mu}_j \quad \text{(Eqn. 3a)}$$

$$R_p = -\frac{\pi E_p}{hc} \sum_{i<j} C_p^{(i)} C_p^{(j)} \boldsymbol{r}_{ij} \cdot (\boldsymbol{\mu}_i \times \boldsymbol{\mu}_j) \quad \text{(Eqn. 3b)}$$

These terms are derived from Fermi's golden rule[22]. An area-normalized Lorentzian lineshape $\mathcal{L}(E - E_p, \gamma)$ with full width at half maximum (FWHM) $\gamma$ is assigned to each transition to account for homogenous broadening.

$$\mathcal{L}(E - E_p, \gamma) = \frac{2}{\pi} \frac{\gamma}{4(E - E_p)^2 + \gamma^2} \quad \text{(Eqn. 4)}$$

The net absorption $A(E)$ and circular dichroism $CD(E)$ spectra are calculated by summing contributions of all the hybrid states.

$$A(E) = \kappa_0 E \sum_p D_p \mathcal{L}(E - E_p, \gamma) \quad \text{(Eqn. 5a)}$$

$$CD(E) = 4\kappa_0 E \sum_p R_p \mathcal{L}(E - E_p, \gamma) \quad \text{(Eqn. 5b)}$$

Where $\kappa_0$ is a constant that accounts for extinction units.

Experimental realization of non-degenerate assemblies

As an experimental testbed for our hypotheses on the chiroptical response of non-degenerate assemblies, we use chiral self-assembled films of CdS magic-size clusters (MSCs)[26,27]. MSCs are closed-shell, atomically identical nanoclusters, where only specific "magic" sizes exist, and cluster sizes between these magic sizes are thermodynamically or kinetically prohibited[28]. MSCs behave more like well-defined molecular semiconductors than variable-size quantum dots; they have discrete, identical transitions[28]. As-synthesized, the CdS MSCs align into helical fibers, with precise nanometer spacing radially[26,27]. Our previous work has shown that meniscus-guided assembly of MSCs can be used to produce chiral superstructure films with a high degree of ordering in all three dimensions. This structuring enables wavelength-scale delocalization of collective chiral excitonic states, resulting in large CD signals not intrinsic to the individual MSCs, with g-factors (CD/Abs) greater than unity[16].

In addition, these MSCs can adopt distinct, discrete transition energies by undergoing a diffusionless, coherent crystalline transformation between two different atomic configurations or 'isomers'. These isomers, α-MSCs and β-MSCs, have primary excitonic transitions at $E_\alpha$=3.82 eV and $E_\beta$=3.97 eV, respectively[29–31]. The α→β isomerization can be achieved by treatment with organic alcohols (e.g., ROH molecules). Exposing chiral α-MSC films to methanol vapor isomerizes the individual MSCs within the films, without disrupting the overall film structure, resulting in a binary component non-degenerate $\alpha_{1-x}\beta_x$ chiroptic structure[29]. The fraction of β-MSCs, $x$, can be tuned via controlling the duration of methanol exposure. This method provides a means to create a tunable chiral assembly with two well-defined, discrete, and interchangeable chromophore states. These chiroptic samples with α-MSCs, β-MSCs, and mixed $\alpha_{1-x}\beta_x$-MSCs serve as an ideal system to study non-degenerate coupling and its resulting chiroptical properties.

To model this material, we treat the chromophores as point dipoles with a single electric transition dipole moment (TDM) and arrange them in a helical lattice (along the z-direction) with hexagonal packing in each plane (x-y plane) (**Fig. 2a**). In the x-y plane, the chromophores are packed in a hexagonal arrangement with a lattice parameter $a_{he}$, which, based on previously reported SAXS results of the experimental system[27], is 3 $nm$. All the chromophore TDMs in a hexagonal plane are aligned in the same direction, within the plane (**Fig. 2a**). The TDMs of chromophores in adjacent z-planes are oriented at a dihedral angle $\theta$ and separated by an axial spacing $a_z$. This structure is a simplified, spatially averaged approximation of the true twisted-fiber hierarchical organization of the MSC assemblies[27], appropriate for comparison with our optical measurements that probe mm-scale regions. The following values, which roughly correlate with the experimental samples, are used for the parameters: $|\boldsymbol{\mu}| = 7\ D, a_{he} = 3\ nm, a_z = 3\ nm, \theta = 1°, \varepsilon = \varepsilon_0$. Note that the transition dipole moments of both MSC species have been approximated as equal. The full simulated structure contains a total of 3,700 chromophores (100 layers with 37 chromophores in each layer). The chromophores in the assembly can be composed of all α-MSCs species (degenerate at $E_\alpha$=3.82 eV) or all β-MSCs species (degenerate at $E_\beta$=3.97 eV) or a mixture of α-MSCs and β-MSCs species (nondegenerate) (**Fig. 2b**).

Using degenerate assembly constructions, the resulting calculated CD and absorption lineshapes closely match the experimental spectra of both the α-MSC and β-MSC degenerate films (**Fig. 2c**), validating our choice of parameters. While the lineshapes are accurately modeled by these simulated structures, due to the limited simulation size, the large experimental g-factors of MSC assemblies are not attainable by the model. We note that MSC assemblies may form either left- or right- handed chiral structures, which exhibit CD of opposite sign and identical absorption. The spectra of both structures are reproduced in the model by reversing the sign of the dihedral angle. In the following analysis, we restrict all our discussions to negative $\theta$ for consistency, without loss of generality.

For non-degenerate assemblies, films composed of $\alpha_{1-x}\beta_x$-MSCs are simulated using two contrasting approaches: Linear Combination (LC) and Full Hamiltonian (FH). The LC approach assumes that the chiroptical properties are additive and completely described by a weighted sum of the spectra of degenerate α-MSC and β-MSC assemblies:

$$CD_{LC}(x, E) = (1 - x)CD_\alpha(E) + xCD_\beta(E) \qquad \text{(Eqn. 6a)}$$
$$A_{LC}(x, E) = (1 - x)A_\alpha(E) + xA_\beta(E) \qquad \text{(Eqn. 6b)}$$

As we will see, lineshapes from the LC method do not account for emergent properties that arise from non-degenerate coupling.

The Full Hamiltonian (FH) approach is more rigorous, explicitly incorporating the non-degeneracy of states into the Hamiltonian in **Equation 2a**, allowing its consequences to arise naturally from the underlying physics. This method implicitly accounts for emergent effects that arise from the non-degeneracy of states. We assume that the alcohol-induced transformation of α-MSC films is not diffusion-limited, such that α→β isomerization takes place uniformly throughout the film, and not just at the surface. Based on this assumption, we model the $\alpha_{1-x}\beta_x$ films as randomly mixed, assigning each chromophore in the structure to be either α-MSCs or β-MSCs (**Fig. 2d**), with a fixed probability. This randomization is mathematically implemented by treating the diagonal elements in the Hamiltonian (**Eqn. 2**) as a Bernoulli random variable, formulated as:

$$E_i = \begin{cases} E_\alpha, & \text{with probability } 1 - x \\ E_\beta, & \text{with probability } \quad x \end{cases} \qquad \text{(Eqn. 7)}$$

Comparing the two model approaches with the experimental spectra of $\alpha_{1-x}\beta_x$ MSC assemblies allows us to identify and isolate this emergent behavior. Note that both approaches, FH and LC, assume the overall assembly and interactions ($V_{ij}$) remain unchanged with composition.

Experimental realization of non-degenerate chiral assemblies

To experimentally test this construction, we collect CD and optical absorption spectra for an $\alpha_{1-x}\beta_x$ MSC film at different compositions $x$, by varying the duration of the MeOH treatment. The value of $x$ is extracted by fitting the absorption as a superposition of the absorptions of α-MSC and β-MSC films. The black curves in **Figure 2d** show the experimentally measured CD (top panel) and absorption (bottom panel) for the composition $x$ = 0.55. The absorption lineshape is composed of two peaks corresponding to the primary excitonic transitions of the α-MSCs ($E_\alpha$ = 3.82 eV) and β-MSCs ($E_\beta$ = 3.97 eV). The difference in energies (i.e., the 'detuning') value is $\delta E = 0.15$ eV. The CD at this composition is a bisignate centered at the midpoint between these two energies. This bisignate is broadened compared to the degenerate assemblies and possesses a flat region around the zero-crossing energy.

Comparing the experimental spectra with the Linear Combination (**Fig. 2d**, labeled 'LC sim.') and Full Hamiltonian (**Fig. 2d**, labeled 'FH sim.') simulation approaches, and with a linear combination of two experimental spectra of degenerate films (**Fig. 2d**, labeled 'LC exp.') reveals similarities and important differences. For the absorption, the simulation curves closely match the experimental results and are mostly invariant across both approaches (**Fig. 2d**, bottom panels). But for the CD lineshapes, there is a pronounced difference between experimental plots and those generated by the LC and FH approaches (**Fig. 2d**, top panels). CD Lineshapes from the linear combinations (LC), both simulated and experimental, deviate from the nearly pure bisignate of the experimental non-degenerate assembly, most notably by producing an additional undulation at ~3.9 eV. The FH model, in contrast, generates a bisignate that closely matches the experiment without this additional undulation feature. The FH model results in a much better fit, as evidenced by plotting the CD error (experiment - simulation) (**Fig. 2d**, middle panels). To rule out potential errors in the extraction of $x$, the undulation at 3.9 eV is confirmed to appear consistently in the LC simulations across compositions, but not in the FH. Furthermore, the close agreement in absorption between the two approaches validates the use of linear superposition fitting of the absorption for determining the compositional parameter $x$. From this analysis, we conclude that the CD lineshapes of non-degenerate assemblies can accurately be simulated using the FH model.

Next, we probe the optical rotation of MSC films using the Kramers-Kronig (K.K.) transformation to extract the circular birefringence (CB) from the CD lineshapes:

$$CB(E) = \frac{2}{\pi} \wp \int_0^\infty \frac{E' CD(E')}{E'^2 - E^2} dE' \qquad \text{(Eqn. 8)}$$

This relationship holds for passive media and dictates that the CB of a material is uniquely determined by its CD[14]. Bisignate CD spectra (such as the degenerate MSC assemblies) generate a CB with a major peak that aligns with the CD zero crossing and minor dips on either side of the resonance (**Fig. 2e**). In degenerate coupled chiral systems, the region of large CB coincides with the absorption peak. This relationship between CD and CB lineshapes has been verified experimentally for CdS MSC films measured through Mueller Matrix Polarimetry (MMP)[16].

Comparing the extracted lineshapes of CB for the $\alpha_{1-x}\beta_x$ assemblies computed from the CD of the experimental measurement, FH simulation, and the LC simulation, there are clear similarities between the experimental and FH simulations and obvious differences between those two and the LC simulation (**Fig. 2f**). Specifically, the LC CB spectra exhibit a decrease in intensity between the two MSC resonances while the FH and experimental CB both exhibit a nearly flatband response spanning between both resonances. The dip in the LC case arises because the Kramers–Kronig transform of the summed CD bisignates is simply the linear combination of the corresponding CB monosignates, reflecting the linearity of the transform (**Fig. 2f,** green). As the FH and experimental data show a near-constant line between the peak that is absent in the linear combination, we conclude that this emergent feature cannot be explained by the simple superposition of the individual components and arises from non-degenerate coupling. We further confirm this hypothesis by simulating CB at the center energy point (~3.9 eV) across different compositions for both the LC and FH simulation approaches (**Fig. 2f**, inset). The resulting trend for LC is a flat line connecting the endpoints, as expected from a superposition, but the FH trend forms a curve, indicating an emergent effect that has a maximum CB at $x = 0.5$, where the number of non-degenerate interactions is maximized because both chromophores are present in equal quantities. This result demonstrates that the enhanced off-resonant CB arises as a consequence of non-degenerate coupling. Importantly, this CB spectral feature (at ~3.9 eV) aligns with the broadened near-zero CD region and a dip in the absorption. This combination of properties from non-degenerate coupling shows promise for improving optical rotation performance.

Designing non-degenerate structures for pure optical rotation

The results from the α$_{1-x}$β$_x$ MSC system (**Fig. 2**) validate the Full Hamiltonian (FH) model and demonstrate the emergence of circular birefringence (CB) spectral features away from resonance in non-degenerate assemblies, favorable for pure rotation. Building on this experimental testbed, we now extend our theoretical approach to establish design rules for maximizing off-resonant CB in assemblies composed of two generalized non-degenerate chromophores (organic molecules, plasmonic nanoparticles, or semiconductor nanoclusters). The hexagonally packed helical structure (**Fig. 2a**) is retained as the model system for the subsequent analysis with parameters $\mu = 7\,D$, $a_{hex} = a_z = 3\,nm$ and $\gamma = 150\,meV$. The system size of 3,700 chromophores (100 layers of 37 hexagonally packed chromophores) is also retained from the previous sections. Additionally, we set the central energy $E_0 = 3.3\,eV$ for subsequent simulations.

To identify optimal chromophore arrangements for pure rotation, we adopt a pairwise decomposition (PD), where the total CD/CB response is approximated to the first order as the sum of contributions from all $^NC_2$ interacting chromophore pairs in the assembly. This assumption reduces the chiroptical spectra of the complex many-body system into tractable, separable contributions, and has been successfully employed in prior studies of exciton-coupled organic systems [32–34]. Additionally, we demonstrate that this framework successfully reproduces the key chiroptical behaviors observed in assemblies operating in this coupling regime, thereby validating its applicability. Within this framework, the spectral features of any two-component, non-degenerate assembly consisting of chromophores A and B can be approximately decomposed into contributions from three types of dimers: A-A, B-B, and A-B (**Fig. 3a**). The CD and CB spectra can thus be represented as:

$$CD(E) \approx c_{AA}CD_{AA}(E) + c_{BB}CD_{BB}(E) + c_{AB}CD_{AB}(E) \quad \text{(Eqn. 9 a)}$$

$$CB(E) \approx c_{AA}CB_{AA}(E) + c_{BB}CB_{BB}(E) + c_{AB}CB_{AB}(E) \quad \text{(Eqn. 9 b)}$$

Where $c_{AA}$, $c_{BB}$, and $c_{AB}$ are scalar weights that encode the multiplicity of each dimer type. Since the dipole coupling term $V_{ij} \propto |r_{ij}|^{-3}$ (**Eqn. 1**), the contributions $CD_{AA}$, $CD_{BB}$, and $CD_{AB}$ are dominated by the nearest neighbor pair of each dimer type. For a hexagonally stacked helical assembly the predominant dimers correspond to axial nearest neighbors with separation $a_z$ and a dihedral angle $\theta$. The CD and CB magnitude and sign of these dimers vary with $\mu$, $\theta$ and $a_z$ as[19]

$$CB_{dimer}, CD_{dimer} \propto \frac{|\mu|^4 \sin 2\theta}{|a_z|^2} \tag{Eqn. 10}$$

The salient lineshape features of chiral dimers, largely independent of the input parameters, are as follows: degenerate dimers (**Fig. 3a**, A-A and B-B) have single absorption Lorentzians, narrow monosignate CB peaks, and bisignate CD, all centered at the chromophore resonance ($E_{A/B}$). In contrast, the non-degenerate dimers (**Fig. 3a**, A-B) produce two distinct absorption peaks at $E_A$ and $E_B$ with broadband CB spanning between them and CD bisignate centered at $E_0 = (E_A + E_B)/2$. Consequently, non-degenerate chromophores in proximity contribute to a broadband, off-resonant CB response coinciding with low CD. These advantageous features of these dimers are, however, accompanied by a reduction in the CB response relative to their degenerate counterparts. Analytical results[19] show that the magnitude of the rotational strength $R_{AB}$ of the hybrid states, which decides the CB and CD magnitude, decays as energetic detuning increases (**Fig. 3b**), according to:

$$CB_{AB}, CD_{AB} \sim R_{AB} \propto \left(1 + \left(\frac{\delta E}{2V_{AB}}\right)^2\right)^{-\frac{1}{2}} \tag{Eqn. 11}$$

Where $V_{AB}$ is the coupling energy between the chromophores in the dimer. That is, the CD and CB from non-degenerate pairs fall off as $\sim 1/\delta E$ at large detuning ($\delta E > 10 V_{AB}$). This behavior reflects a fundamental trade-off in the nature of the non-degenerate contributions: increasing energetic detuning generates the desired off-resonant CB, but simultaneously weakens the rotational strength.

To scale up this simulation into large assemblies, we create a large structure ($N = 3700$) in which A and B chromophores are randomly mixed, and simulate the optical response using the Full Hamiltonian (FH) approach (**Fig. 3a**, rightmost panel). We first consider $\delta E = 0.6\ eV$, since large detuning separates the absorption peaks, resulting in a spectral window where CB can persist with low loss and CD, desirable for pure rotation. In the randomly mixed assembly, the A-A, B-B and A-B dimer types occur in equal multiplicities, but the spectral contribution of the A-B dimers is relatively weak because of the detuning dependence (**Fig. 3b**, **Eqn. 11**). As a result, the CB and CD spectra of the assembly are dominated by the A-A and B-B contributions, that is, narrow monosignate CB and bisignate CD features centered at both $E_A$ and $E_B$ (**Fig. SX**). Consequently, the desirable broadband off-resonant CB contribution from A-B interactions becomes negligible in randomly mixed structures with large detuning.

This result provides a design objective: to maximize the off-resonant CB, the structure must enhance A-B dimer contributions. To this end, we propose an ABAB stacking configuration (**Fig. 3c**), in which alternating hexagonal planes are populated exclusively by one chromophore type, thereby maximizing axial A-B nearest neighbors. As an initial baseline case, we set $\theta = 45°$. In this configuration, the non-degenerate A-B nearest neighbors are oriented at $45°$, for which $\sin 2\theta$ (**Eqn. 10**) is maximized, thereby maximizing their contribution. At the same time, the nearest degenerate neighbors (A-A and B-B), which occur between alternating planes, are oriented at $90°$, setting $\sin 2\theta = 0$ and thus eliminating their CD and CB contributions. This choice of geometry is therefore expected to selectively enhance the desired non-degenerate response while extinguishing the degenerate background.

We then simulate the spectral responses of the proposed 45° ABAB structure with fully degenerate assemblies and randomly mixed non-degenerate assemblies, and evaluate their respective performance for achieving pure optical rotation (**Figs. 3d, 3e**). For consistency, we set $\theta = 45°$ for both the degenerate and randomly mixed assemblies.

The degenerate chiral assemblies (**Fig. 3d**) exhibit the fundamental features in the CD, CB, and Abs observed in the degenerate dimers. Chiral assemblies of this type with long-range periodic order have been shown to possess large CD (g-factors >1), which result in large CB values and large optical rotation. But such a material is unfit for applications because the CB active region (**Fig. 3d**, gray area) aligns with the absorption peak, leading to large losses, and the large circular dichroism, which is also in the proximity of the CB (**Fig. 3d**, gray area), leads to degradation of the linear polarization into an elliptical signal (**Fig. 3d**, bottom).

For the nondegenerate assemblies (with $\delta E = 0.6$ eV), the two chromophore arrangement modes (randomly mixed and ABAB) lead to very different spectra (**Fig. 3e**). While the randomly mixed structure (**Fig. 3e**, pink curves) is dominated by the degenerate A-A and B-B contributions as discussed earlier, the proposed ABAB configuration (**Fig. 3e**, black curves) exhibits a CB and CD lineshape closely resembling that of an isolated A-B dimer (**Fig. 3d**). Specifically, the spectrum displays a strong broadband CB feature spanning the region between $E_A$ and $E_B$, accompanied by a bisignate CD centered at $E_0 = (E_A + E_B)/2$, consistent with the behavior predicted by the pairwise decomposition analysis. Importantly, the off-resonant CB of the ABAB structure greatly exceeds that of the randomly mixed assembly with the same geometric parameters, even though the absorption spectra of the two structures are nearly identical. This result demonstrates that the enhanced CB originates from the controlled arrangement of chromophores rather than differences in oscillator strengths. These results validate our design strategy of enforcing an ABAB chromophore arrangement to selectively amplify non-degenerate contributions and achieve a broadband, low-loss CB response suitable for pure optical rotation.

We simulate the variation of the CB, CD, and absorption of the proposed ABAB structure with $\theta = 45°$ as a function of chromophore detuning $\delta E$ (**Fig. 4a**), which reveals a fundamental trade-off governing this system. As discussed earlier, large values of $\delta E$ with respect to the resonance linewidth creates a window around $E_0 = (E_A + E_B)/2$ (gray dotted line) with low CD (**Fig. 4a**, middle panel) and absorption (**Fig. 4a**, bottom panel). However, the CB in the same spectral range (**Fig. 4a**, top panel) sharply decreases with increasing $\delta E$. Consequently, the CB magnitude and the width of the low-CD and low-absorption window cannot be increased independently. This trade-off arises from the weakening of rotational strength in non-degenerate assemblies (**Eqn. 11**) and constitutes a critical design constraint.

Among the assembly parameters, $\mu$, $a_z$, and $a_{hex}$ do not significantly alter the spectral lineshapes, contributing primarily as overall scaling factors in this regime of coupling strength. In contrast, the CD and CB lineshapes vary strongly with the inter-layer dihedral angle $\theta$, which controls the relative coupling strengths between degenerate (A-A, B-B) and non-degenerate (A-B) layers. To examine this dependence, we simulate the chiroptical response of the ABAB assembly for $\theta = 30°$ and $60°$ (**Fig. 4b**) at $\delta E = 0.6$ eV and compare it with the baseline $\theta = 45°$ case discussed earlier. While the 45° configuration produced spectra dominated by the non-degenerate A-B interaction and therefore resembled the simple A-B dimer features (**Fig. 3c**), deviations from this angle introduce additional contributions from the degenerate A-A and B-B pairs.

These three angles represent three general regimes: $\theta < 45°$, $\theta = 45°$, and $\theta > 45°$, with each regime exhibiting similar spectral behavior [**SI**]. In the 45° case, the degenerate A-A and B-B contributions vanish because $\sin 2\theta = 0$, resulting in CB and CD dominated by the non-degenerate A-B dimer features, as discussed earlier. For $\theta < 45°$, however, the degenerate contributions interfere destructively with the non-degenerate A–B contribution, reducing the CB magnitude around $E_0$ (gray dotted line), as seen in the 30° case (**Fig. 4b**, 30° in top panel). In contrast, for $\theta > 45°$, the degenerate interactions interfere constructively with the non-degenerate response, leading to an enhanced CB magnitude (**Fig. 4b**, 60° in top panel). The CD magnitude follows the same trend, while the absorption lineshapes remain unchanged with angle. Additional angles and the full pairwise decomposition analysis as a function of $\theta$ are provided in the Supporting Information.

To examine the combined influence of chromophore detuning $\delta E$ and inter-layer angle $\theta$ on pure optical rotation performance, we map the chiroptical response of the ABAB structure across this parameter space. We simulate the chiroptical response (**Fig. 4c**) as a function of detuning $\delta E$ for the same three representative angles discussed earlier ($\theta = 30°, 45°, 60°$). Against the common horizontal axis of $\delta E$, we plot the CB at $E_0$ (here, 3.3 eV) (top panel), the CD slightly offset from the $E_0$ zero-crossing (3.35 eV) (middle panel), and the ratio CB/Abs at $E_0$ (bottom panel), which serves as a metric for pure optical rotation by quantifying rotation relative to absorption loss.

The resulting trends reveal two distinct regimes. At very small detunings, spanning the transition from degeneracy to approximately $\delta E \approx 0.1$ eV, comparable to the resonance linewidth $\gamma$, the spectral response exhibits a transient variation. For nearly degenerate structures (small $\delta E$), the CB, CD, and CB/Abs are largest for smaller angles. However, this ordering reverses at larger detunings (**Fig. 4 c, and 4c insets**), with larger angles possessing greater values for CB, CD, and CB/Abs. Importantly, the CD decreases with detuning more rapidly than the CB. Additionally, in the $\theta > 45°$ range, the CB/Abs ratio increases with detuning, providing a route to optimize pure optical rotation. As the polarization quality improves at larger detuning due to reduced loss and CD, the sample thickness can be increased simultaneously to compensate for the reduction in CB magnitude. The detailed strategy for achieving pure rotation with the ABAB architecture by balancing detuning and thickness is presented in the following section. The analysis in **Figure 4c**, therefore, identifies $\theta > 45°$ as a particularly favorable regime for achieving pure optical rotation. Notably, this anomalous scaling of CB/Abs with detuning arises from the collective behavior of the three-dimensional assembly rather than the response of isolated dimers, highlighting the importance of collective excitonic interactions in determining the chiroptical response of these structures.

So far, we have only discussed the CB, CD, and absorption lineshapes of our proposed ABAB geometry, without considering their magnitudes and quantitative transformation of linearly polarized light. In the following section we use the Stokes-Mueller formalism to estimate the pure optical rotation performance of this system with the following key assumptions: Since our model does not account for absolute magnitudes of the optical effects, we externally assign the peak g-factor ($g_0 = \max(CD/Abs)$) and peak absorbance ($A_0$) values to the degenerate assemblies based on previous experimental results of the CdS MSC system[16]. We assume that the assemblies can achieve a peak g-factor at degeneracy $g_0(\delta E = 0) = g_{\text{deg}} = 1$ (as shown previously for MSC assemblies[16]). To simplify the analysis, we further assume that this peak value has been achieved for a degenerate structure, regardless of the assembly parameters ($\mu, a_z, a_{hex}$). Although this idealization likely represents an upper bound, it enables us to isolate the effect of energetic detuning on the optical response. The dependence of the g-factor on detuning $g_0(\delta E)$ is inherited from the coupled-oscillator model, following the dynamics that manifest in **Equation 10**. The peak absorption for the assembly at degeneracy $A_0(\delta E = 0) = A_{\text{deg}}$, directly corresponds to the sample thickness. Since this analysis pertains to solution-processed materials, thickness, and by extension, $A_{\text{deg}}$ can be scaled freely. Like the g-factor, the variation of the peak sample absorbance with detuning $A_0(\delta E)$ is also dictated by the coupled-oscillator model. Note that we report $A_0$ in base-10 Beer-Lambert convention as reported on most UV-Vis spectrometers (defined as $A = \log_{10}\left(\frac{I_0}{I}\right)$, where $I$ and $I_0$ are the intensities of the transmitted and incident signals). To benchmark the optical rotation performance of the proposed ABAB non-degenerate assemblies, we calculate the optical rotation (OR, Ψ), ellipticity (χ), and transmission (T) using the assumed g-factor and sample absorbance values within the Stokes-Mueller approach. $\psi$, $\chi$, and T are intimately tied and bear resemblance with the CB, CD, and absorption, respectively (**Fig. 5b**). We confirm that this method correctly calculates the magnitudes of rotation, ellipticity, and transmission by comparing modelling results to Mueller Matrix Polarimetry measurements of an experimentally measured α-MSC film (degenerate case).

To evaluate the achievable optical rotation, we apply this framework to the design strategy developed in the previous section (**Fig. 4c**), namely, simultaneously increasing the chromophore detuning ($\delta E$) and sample thickness (parameterized by $A_{\text{deg}}$) to maintain a large CB magnitude while improving the quality of the transmitted polarization. We therefore calculate the optical rotation ($\psi$), ellipticity ($\chi$), and transmission ($T$) for the ABAB structure with $\theta = 60°$ over a $\delta E$–$A_{\text{deg}}$ parameter space (**Fig. 5c,d**). As illustrated earlier (**Fig. 3e**), the lineshapes show promise for pure optical rotation in the region around $E = E_0$, the midpoint between the two chromophore resonances. To quantify this performance, we define the following figures of merit (FOMs):
- Optical rotation ($\psi$) at $E_0$,
- Bandwidth of the region around $E_0$ with ellipticity lower than 1° ($w_\chi$, denoted by gray region in the middle panel of **Fig. 4b**) and
- The total transmission at $E_0$ ($T$).

Colormaps of these three FOMs (**Fig. 5c**) illustrate the variation of $\psi$(top), $w_\chi$(middle), and $T$(bottom) across the $\delta E$–$A_{\text{deg}}$ parameter space and provide a guide for designing such systems. The optical rotation $\psi$ (**Fig. 5c**, top panel) increases with $A_{\text{deg}}$(equivalently, film thickness) and decreases with increasing detuning $\delta E$, reflecting the reduction of rotational strength at large detuning (**Eqn. 11**). In contrast, the low-ellipticity bandwidth $w_\chi$ (**Fig. 5c**, middle panel) increases with $\delta E$ due to both the separation of the CD resonances and the weakening of rotational strength, but decreases with increasing thickness due to the accumulation of CD. The transmission $T$ (**Fig. 5c**, bottom panel) increases with detuning as the absorption resonances move away from $E_0$. To identify practically useful operating points, we select target values for the three FOMs: an optical rotation of $\psi = 20°$ at $E_0$ (**Fig. 5c**, red line in top panel), a low-ellipticity bandwidth $w_\chi = 50$ meV (12 THz) (**Fig. 5c**, blue line in middle panel), and a transmission $T = 40\%$ at $E_0$ (**Fig. 5c**, beige line in bottom panel).

These three contour lines are then combined (**Fig. 5d**) to illustrate how the three FOMs vary jointly with detuning and sample absorbance. The red contour corresponds to $\psi = 20°$, with the region above this curve satisfying $\psi > 20°$. Similarly, the region below the yellow contour corresponds to transmission $T > 40\%$, while the region below the blue contour corresponds to a low-ellipticity bandwidth $w_\chi > 50$ meV. Overall, the improvement in optical rotation brought on by increasing the sample thickness outweighs the deterioration of polarization quality, in conjunction with **Figure 4c**. This result is evidenced by the fact that the $w_\chi = 50\ meV$ contour (**Fig. 5d**, blue curve) crosses the $\psi = 20°$ contour (**Fig. 5d**, red curve) at larger detuning and larger thickness values, resulting in the formation of a unique and highly desirable region in the parameter space where all three FOMs are optimized (**Fig. 5d**, green region). This region defines the optimal operating regime for achieving strong optical rotation with low ellipticity and moderate transmission in the proposed ABAB assemblies.

**Outlook**

The pure rotation performance — simultaneous optimization of optical rotation ($\psi > 20°$) and polarization quality ($w_\chi > 50 meV$, $T > 40\%$) — achieved by our theoretical system is competitive with pure optical rotation metamaterials in the literature[3,6–13]. The advantage of these MSC materials is that, unlike lithographic metamaterials, the colloids can be solution-processed and can exhibit strong optical activity at shorter wavelengths, including in the UV, as already demonstrated with CdS MSCs[16].

The region in parameter space that presents pure rotation of linear polarizations in our simulations is achieved by creating a structure with a peak absorbance at degeneracy ($A_\text{deg}$) of 5 and with two chromophores detuned by ~ 1 eV. As a materials example, CdS MSCs processed into films[16] that are 1 - 1.5 microns thick have similar chromophore densities and oscillator strengths to those used in this model, and exhibit an absorbance of ~1. Therefore, achieving the target FOM, which requires an absorbance of ~5, should be possible with films only a few microns thick. In contrast, off-the-shelf quartz components that perform as well in magnitude of rotation are at millimeter-scale thicknesses. The detuning parameters are also within reach: chromophore resonance detunings greater than 1 eV are feasible within cadmium chalcogenide MSCs themselves: CdTe MSCs have a resonance at 2.75 eV (450 nm)[16], which is detuned by 1.22 eV with respect to the β-isomer of CdS MSCs. The structure considered in this study is a simplified hexagonally packed helix with parallel dipoles within each layer, resulting in zero in-plane coupling. Our results in **Figure 4** indicate that degenerate interactions can be leveraged to enhance the magnitude of the circular birefringence (CB). More complex assembly architectures that exploit in-plane degenerate interactions and arrange chromophores to constructively enhance CB could further improve the optical response. Additionally, with the parameters used in this study ($\mu = 7\ D$, $a_z = 3\ nm$), the coupling strength is of the order of 1 meV. Increasing this coupling strength would slow the decay of rotational strength with detuning (**Eqn. 11**), resulting in larger CB magnitudes at higher detuning and enabling improved performance of the proposed system.

In summary, we have shown that non-degeneracy of chromophore excited states offers a powerful and previously unexplored route to achieving pure optical rotation in self-assembled materials. We used a coupled-oscillator formalism and demonstrated that detuning between chromophores can enlarge the separation of circularly dichroic resonances and create a broad, low-loss circular birefringence window analogous to that exploited in engineered metamaterials. Mixed assemblies of magic-sized nanocluster isomers confirm this mechanism and validate the predictive capability of the model. Building on this insight, we identified a simple ABAB stacking motif that enhances pure rotation performance and enables simultaneous realization of large polarization rotation, minimal ellipticity, and appreciable transmission—figures of merit that are typically inaccessible in self-assembled chiral systems.

Because our strategy relies only on the coupling between chromophores with distinct transition energies, it is broadly generalizable across molecular, semiconductor, and plasmonic building blocks with dipole-allowed transitions. Our work establishes non-degenerate coupling as a scalable design principle for generating pure optical rotation in solution-processed materials and expands the landscape of architectures capable of producing low-loss, polarization-preserving chiroptical functionality. We anticipate that this approach will facilitate the development of new chiral photonic platforms for sensing, polarization control, and integrated optical technologies.

**Acknowledgements**
Funding: This work was supported in part by the National Science Foundation (NSF) under award nos. CMMI-2120947 and DMR-2344586. This work made use of the Cornell Center for Materials Research shared instrumentation facility. Mueller Matrix Polarimetry (MMP) data was collected at the Diamond Light Source B23 beamline under proposal SM34669. We acknowledge support from the Kavli Institute at Cornell for Nanoscale Science (KIC).

**References**

(1) Oh, S. S.; Hess, O. Chiral Metamaterials: Enhancement and Control of Optical Activity and Circular Dichroism. *Nano Convergence* **2015**, *2* (1), 24. https://doi.org/10.1186/s40580-015-0058-2.
(2) Deng, Q.-M.; Li, X.; Hu, M.-X.; Li, F.-J.; Li, X.; Deng, Z.-L. Advances on Broadband and Resonant Chiral Metasurfaces. *npj Nanophoton.* **2024**, *1* (1), 20. https://doi.org/10.1038/s44310-024-00018-5.
(3) Katsantonis, I.; Manousidaki, M.; Koulouklidis, A. D.; Daskalaki, C.; Spanos, I.; Kerantzopoulos, C.; Tasolamprou, A. C.; Soukoulis, C. M.; Economou, E. N.; Tzortzakis, S.; Farsari, M.; Kafesaki, M. Strong and Broadband Pure Optical Activity in 3D Printed THz Chiral Metamaterials. *Advanced Optical Materials* **2023**, *11* (18), 2300238. https://doi.org/10.1002/adom.202300238.
(4) Born, M.; Wolf, E. *Principles of Optics: Electromagnetic Theory of Propagation, Interference and Diffraction of Light*, Seventh (expanded) edition, 13th printing.; Cambridge University Press: Cambridge, 2017.
(5) Nye, J. F. *Physical Properties of Crystals: Their Representation by Tensors and Matrices*, Reprinted.; Oxford science publications; Clarendon Press: Oxford, 2012.
(6) Rogacheva, A. V.; Fedotov, V. A.; Schwanecke, A. S.; Zheludev, N. I. Giant Gyrotropy Due to Electromagnetic-Field Coupling in a Bilayered Chiral Structure. *Phys. Rev. Lett.* **2006**, *97* (17), 177401. https://doi.org/10.1103/PhysRevLett.97.177401.

(7) Song, K.; Ding, C.; Su, Z.; Liu, Y.; Luo, C.; Zhao, X.; Bhattarai, K.; Zhou, J. Planar Composite Chiral Metamaterial with Broadband Dispersionless Polarization Rotation and High Transmission. *J. Appl. Phys.* **2016**, *120* (24), 245102. https://doi.org/10.1063/1.4972977.

(8) Decker, M.; Ruther, M.; Kriegler, C. E.; Zhou, J.; Soukoulis, C. M.; Linden, S.; Wegener, M. Strong Optical Activity from Twisted-Cross Photonic Metamaterials. *Opt. Lett., OL* **2009**, *34* (16), 2501–2503. https://doi.org/10.1364/OL.34.002501.

(9) Hannam, K.; Powell, D. A.; Shadrivov, I. V.; Kivshar, Y. S. Dispersionless Optical Activity in Metamaterials. *Applied Physics Letters* **2013**, *102* (20), 201121. https://doi.org/10.1063/1.4807438.

(10) Li, Y.-R.; Hung, Y.-C. Dispersion-Free Broadband Optical Polarization Rotation Based on Helix Photonic Metamaterials. *Opt. Express* **2015**, *23* (13), 16772. https://doi.org/10.1364/OE.23.016772.

(11) Barr, L. E.; Díaz-Rubio, A.; Tremain, B.; Carbonell, J.; Sánchez-Dehesa, J.; Hendry, E.; Hibbins, A. P. On the Origin of Pure Optical Rotation in Twisted-Cross Metamaterials. *Sci Rep* **2016**, *6* (1), 30307. https://doi.org/10.1038/srep30307.

(12) Li, Z.; Caglayan, H.; Colak, E.; Zhou, J.; Soukoulis, C. M.; Ozbay, E. Coupling Effect between Two Adjacent Chiral Structure Layers. *Opt. Express* **2010**, *18* (6), 5375. https://doi.org/10.1364/OE.18.005375.

(13) Kim, T.-T.; Oh, S. S.; Park, H.-S.; Zhao, R.; Kim, S.-H.; Choi, W.; Min, B.; Hess, O. Optical Activity Enhanced by Strong Inter-Molecular Coupling in Planar Chiral Metamaterials. *Sci Rep* **2014**, *4* (1), 5864. https://doi.org/10.1038/srep05864.

(14) Polavarapu, P. L. Kramers- Kronig Transformation for Optical Rotatory Dispersion Studies. *The Journal of Physical Chemistry A* **2005**, *109* (32), 7013–7023.

(15) Schulz, M.; Zablocki, J.; Abdullaeva, O. S.; Brück, S.; Balzer, F.; Lützen, A.; Arteaga, O.; Schiek, M. Giant Intrinsic Circular Dichroism of Prolinol-Derived Squaraine Thin Films. *Nature communications* **2018**, *9* (1), 2413.

(16) Ugras, T. J.; Carson, R. B.; Lynch, R. P.; Li, H.; Yao, Y.; Cupellini, L.; Page, K. A.; Wang, D.; Arbe, A.; Bals, S.; Smieska, L.; Woll, A. R.; Arteaga, O.; Jávorfi, T.; Siligardi, G.; Pescitelli, G.; Weinstein, S. J.; Robinson, R. D. Transforming Achiral Semiconductors into Chiral Domains with Exceptional Circular Dichroism. *Science* **2025**, *387* (6733), eado7201. https://doi.org/10.1126/science.ado7201.

(17) Trazo, J. G.; Serrano Freijeiro, A.; Mychinko, M.; Kutalia, N.; Impéror-Clerc, M.; Bals, S.; Liz-Marzán, L. M.; Pérez-Juste, J.; Pastoriza-Santos, I.; Hamon, C. Tunable Plasmonic Circular Dichroism of Hierarchical Chiral Assemblies. *Chem. Mater.* **2025**, *37* (16), 6237–6245. https://doi.org/10.1021/acs.chemmater.5c00909.

(18) Lu, J.; Xue, Y.; Bernardino, K.; Zhang, N.-N.; Gomes, W. R.; Ramesar, N. S.; Liu, S.; Hu, Z.; Sun, T.; de Moura, A. F.; others. Enhanced Optical Asymmetry in Supramolecular Chiroplasmonic Assemblies with Long-Range Order. *Science* **2021**, *371* (6536), 1368–1374.

(19) Harada, N.; Nakanishi, K. Circular Dichroic Spectroscopy: Exciton Coupling in Organic Stereochemistry. **1983**.

(20) Kim, M.-H.; Ulibarri, L.; Keller, D.; Maestre, M. F.; Bustamante, C. The Psi-Type Circular Dichroism of Large Molecular Aggregates. III. Calculations. *The Journal of Chemical Physics* **1986**, *84* (6), 2981–2989. https://doi.org/10.1063/1.450279.

(21) Wade, J.; Hilfiker, J. N.; Brandt, J. R.; Liirò-Peluso, L.; Wan, L.; Shi, X.; Salerno, F.; Ryan, S. T.; Schöche, S.; Arteaga, O.; others. Natural Optical Activity as the Origin of the Large Chiroptical Properties in π-Conjugated Polymer Thin Films. *Nature communications* **2020**, *11* (1), 6137.

(22)     Andrushchenko, V.; Bouř, P. Circular Dichroism Enhancement in Large DNA Aggregates Simulated by a Generalized Oscillator Model. *J Comput Chem* **2008**, *29* (16), 2693–2703. https://doi.org/10.1002/jcc.21015.
(23)     Hao, C.; Xu, L.; Ma, W.; Wang, L.; Kuang, H.; Xu, C. Assembled Plasmonic Asymmetric Heterodimers with Tailorable Chiroptical Response. *Small* **2014**, *10* (9), 1805–1812. https://doi.org/10.1002/smll.201303755.
(24)     Pescitelli, G. ECD Exciton Chirality Method Today: A Modern Tool for Determining Absolute Configurations. *Chirality* **2022**, *34* (2), 333–363. https://doi.org/10.1002/chir.23393.
(25)     Auguié, B.; Alonso-Gómez, J. L.; Guerrero-Martínez, A.; Liz-Marzán, L. M. Fingers Crossed: Optical Activity of a Chiral Dimer of Plasmonic Nanorods. *The Journal of Physical Chemistry Letters* **2011**, *2* (8), 846–851.
(26)     Nevers, D. R.; Williamson, C. B.; Savitzky, B. H.; Hadar, I.; Banin, U.; Kourkoutis, L. F.; Hanrath, T.; Robinson, R. D. Mesophase Formation Stabilizes High-Purity Magic-Sized Clusters. *J. Am. Chem. Soc.* **2018**, *140* (10), 3652–3662. https://doi.org/10.1021/jacs.7b12175.
(27)     Han, H.; Kallakuri, S.; Yao, Y.; Williamson, C. B.; Nevers, D. R.; Savitzky, B. H.; Skye, R. S.; Xu, M.; Voznyy, O.; Dshemuchadse, J.; Kourkoutis, L. F.; Weinstein, S. J.; Hanrath, T.; Robinson, R. D. Multiscale Hierarchical Structures from a Nanocluster Mesophase. *Nat. Mater.* **2022**, *21* (5), 518–525. https://doi.org/10.1038/s41563-022-01223-3.
(28)     Pun, A. B.; Mazzotti, S.; Mule, A. S.; Norris, D. J. Understanding Discrete Growth in Semiconductor Nanocrystals: Nanoplatelets and Magic-Sized Clusters. *Accounts of chemical research* **2021**, *54* (7), 1545–1554.
(29)     Williamson, C. B.; Nevers, D. R.; Nelson, A.; Hadar, I.; Banin, U.; Hanrath, T.; Robinson, R. D. Chemically Reversible Isomerization of Inorganic Clusters. *Science* **2019**, *363* (6428), 731–735. https://doi.org/10.1126/science.aau9464.
(30)     Lynch, R. P.; Ugras, T. J.; Robinson, R. D. Discovery of Isomerization Intermediates in CdS Magic-Size Clusters. *ACS Nano* **2024**, acsnano.4c08319. https://doi.org/10.1021/acsnano.4c08319.
(31)     Lynch, R. P.; Robinson, R. D. Isomerization of Inorganic Nanomaterials. *MRS Communications* **2025**, 1–20.
(32)     Wiesler, W. T.; Vazquez, J. T.; Nakanishi, K. Pairwise Additivity in Exciton-Coupled CD Curves of Multichromophoric Systems. *J. Am. Chem. Soc.* **1987**, *109* (19), 5586–5592. https://doi.org/10.1021/ja00253a005.
(33)     Liu, H. W.; Nakanishi, K. Pyranose Benzoates: An Additivity Relation in the Amplitudes of Exciton-Split CD Curves. *J. Am. Chem. Soc.* **1982**, *104* (5), 1178–1185. https://doi.org/10.1021/ja00369a004.
(34)     Nehira, T.; Parish, C. A.; Jockusch, S.; Turro, N. J.; Nakanishi, K.; Berova, N. Fluorescence-Detected Exciton-Coupled Circular Dichroism:  Scope and Limitation in Structural Studies of Organic Molecules. *J. Am. Chem. Soc.* **1999**, *121* (38), 8681–8691. https://doi.org/10.1021/ja990936b.
(35)     Yao, Y.; Ugras, T. J.; Meyer, T.; Dykes, M.; Wang, D.; Arbe, A.; Bals, S.; Kahr, B.; Robinson, R. D. Extracting Pure Circular Dichroism from Hierarchically Structured CdS Magic Cluster Films. *ACS Nano* **2022**, *16* (12), 20457–20469. https://doi.org/10.1021/acsnano.2c06730.
(36)     Stokes, G. G. On the Composition and Resolution of Streams of Polarized Light from Different Sources. *Transactions of the Cambridge Philosophical Society* **1851**, *9*, 399.

**Figure 1**

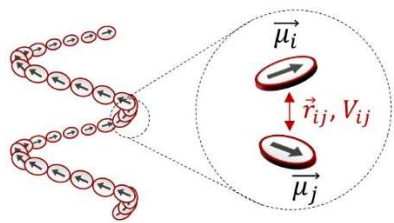
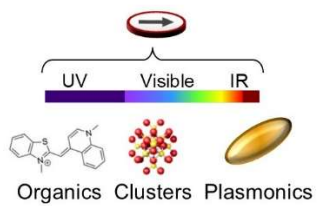
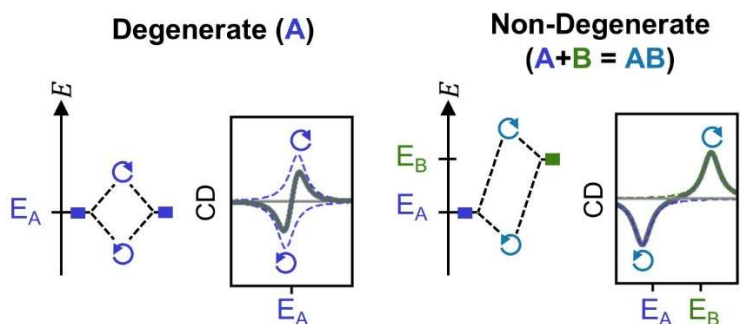
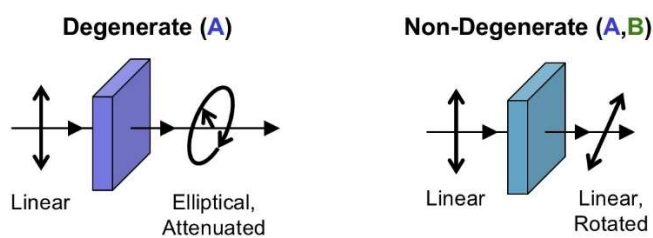

**Figure 1**: Self-assembled chiral superstructures. a) Schematic of coupling between the transition dipole moments ($\mu_i$, arrows indicate direction) of the constituent chromophores. Dipole-dipole interactions ($V_{ij}$) lead to the formation of large delocalized collective chiral states. b) Such assemblies have been experimentally realized with organic, semiconductor nanocrystal, and plasmonic nanoparticle building blocks, across the electromagnetic spectrum. c) The collective chiral states have their resonances split from the resonance of the building block chromophore ($E_A$), and interact differently with left-and right-handed circularly polarized light, leading to a bisignate circular dichroism (CD). In the case where chromophores are non-degenerate ($E_A$ and $E_B$) with large detuning, the resulting CD peaks have a more prominent separation. d) Degenerate assemblies are not suitable for optical rotation applications as their circular birefringence (CB) overlaps with the resonances. In this work, we demonstrate that creating chiral assemblies of non-degenerate chromophores is a viable strategy to realize pure optical rotation in scalable, solution-processed systems.

**Figure 2**

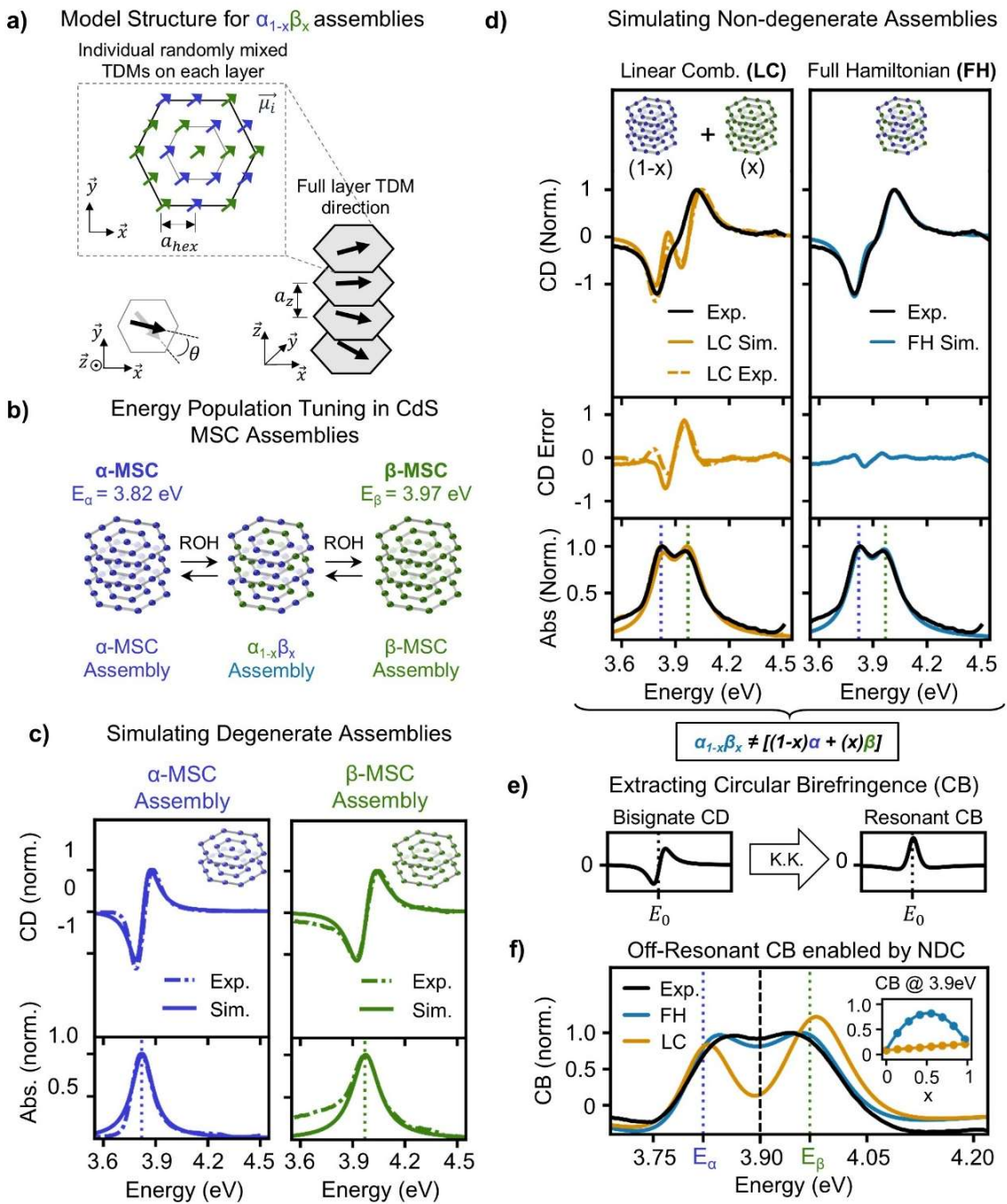

Figure 2: Employing $\alpha_{1-x}\beta_x$ CdS magic-sized cluster (MSC) films to validate the coupled-oscillator model and demonstrate that off-resonant CB contributions emerge from non-degenerate coupling. a) Schematic of hexagonally packed helical lattice used as the simulation geometry in this study. b) Schematic of $\alpha_{1-x}\beta_x$ MSC films, whose building block chromophores have switchable energy states ($E_\alpha$ and $E_\beta$), allowing the realization of a non-degenerate chiral assembly with tunable composition. c) Experimental validation of the model against degenerate MSC films, the control cases. c) Comparison of the experimental CD and absorption (black curves) with Linear Combination (LC, left column) and Full Hamiltonian (FH, right column) simulations. The middle panels plot the difference between the experimental and simulated lineshapes. e) Schematic of Kramers-Kronig (K.K.) transform, used to extract circular birefringence (CB) from CD. f) Comparing CB extracted from experiments with FH and LC simulations. Inset: CB at 3.9 eV simulated across compositions (x), for FH and LC simulations.

**Figure 3**

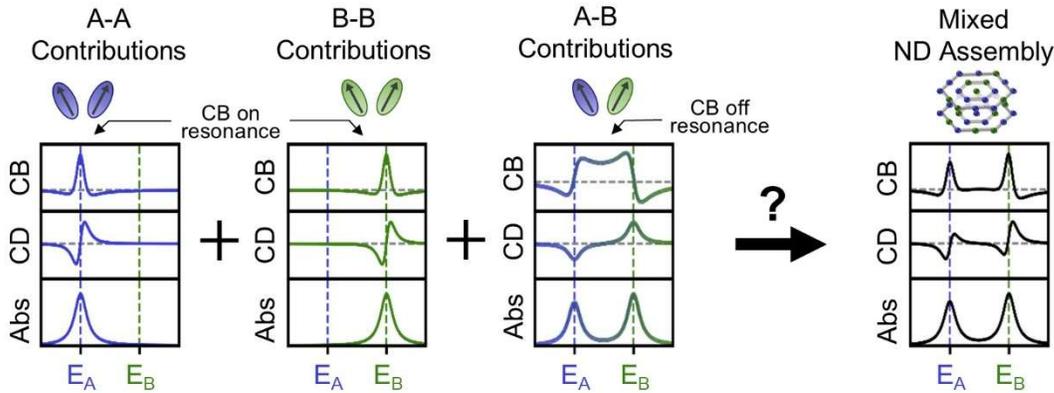

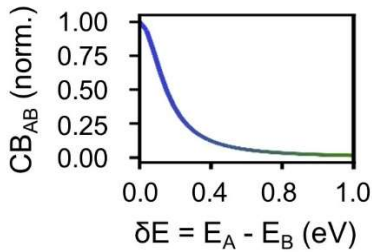

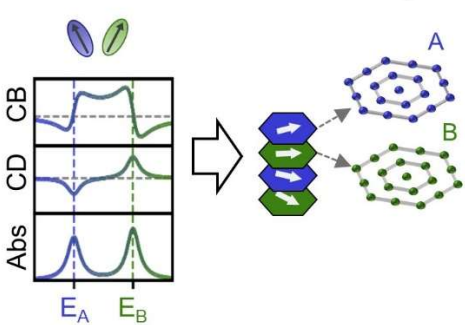

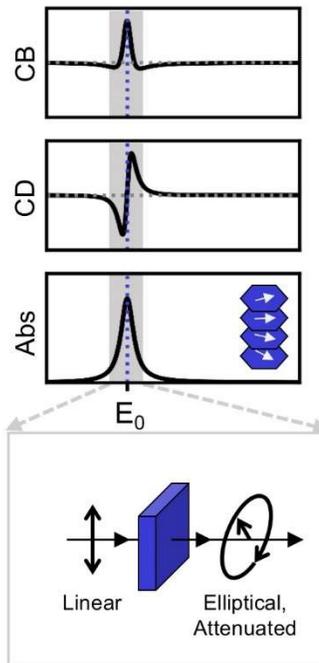

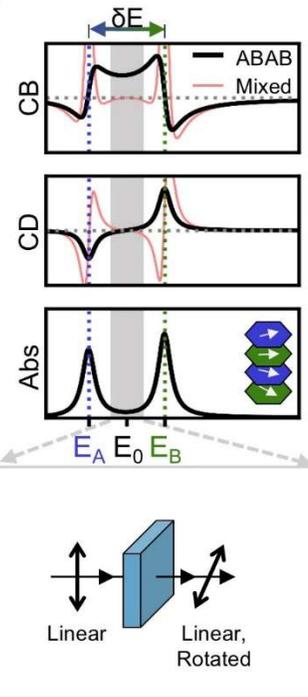

Figure 3: a) Pairwise decomposition approximation of the chiroptical response of AB non-degenerate assemblies into A-A, B-B, and A-B contributions. The desirable properties — broadband off-resonant CB and a low-CD window — arise solely from non-degenerate coupling, motivating the design of ABAB structures that maximize these interactions. b) Decay of A-B contributions with detuning $\delta E$ c) Proposed ABAB layered structure to maximize A-B contributions d), e) Comparison between the CB, CD, and Absorption lineshapes for a degenerate structure (d) and an optimized ABAB layered non-degenerate heterostructure, where the resonances of A and B are detuned by δE (e) Non-degenerate structures simultaneously possess low absorption, CD, and high CB in the gray region, allowing for pure optical rotation.

**Figure 4**

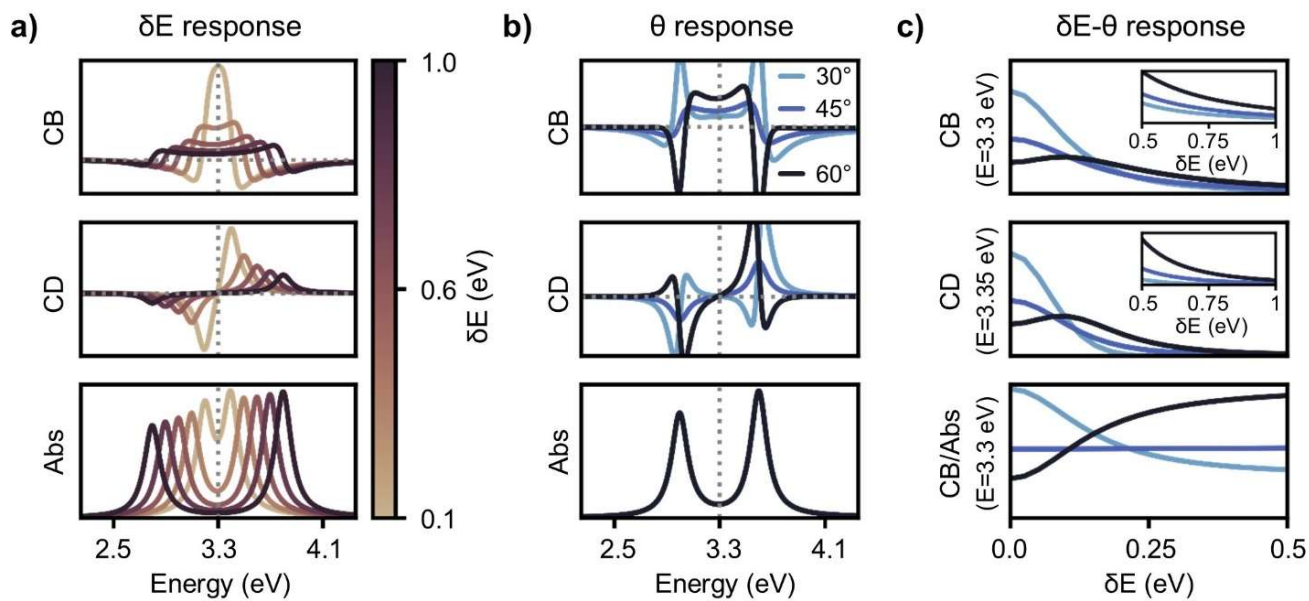

Figure 4: Parametric variation of chromophore detuning $\delta E$ and dihedral angle $\theta$ in the ABAB system. a) Variation of CB, CD and absorption with chromophore detuning. b) CB, CD and Abs lineshapes for an ABAB structure with $\delta E = 0.6\ eV$ at $\theta = 30°, 45°, 90°$ c) CB (at $E = E_0 = 3.3\ eV$, midpoint energy between resonances), CD (at $E = 3.35\ eV$, slightly offset from midpoint) and absorption (at $E = 3.3 eV$) varied with detuning. The insets show the variation of the corresponding parameters at large detuning (0.5-1 eV).

**Figure 5**

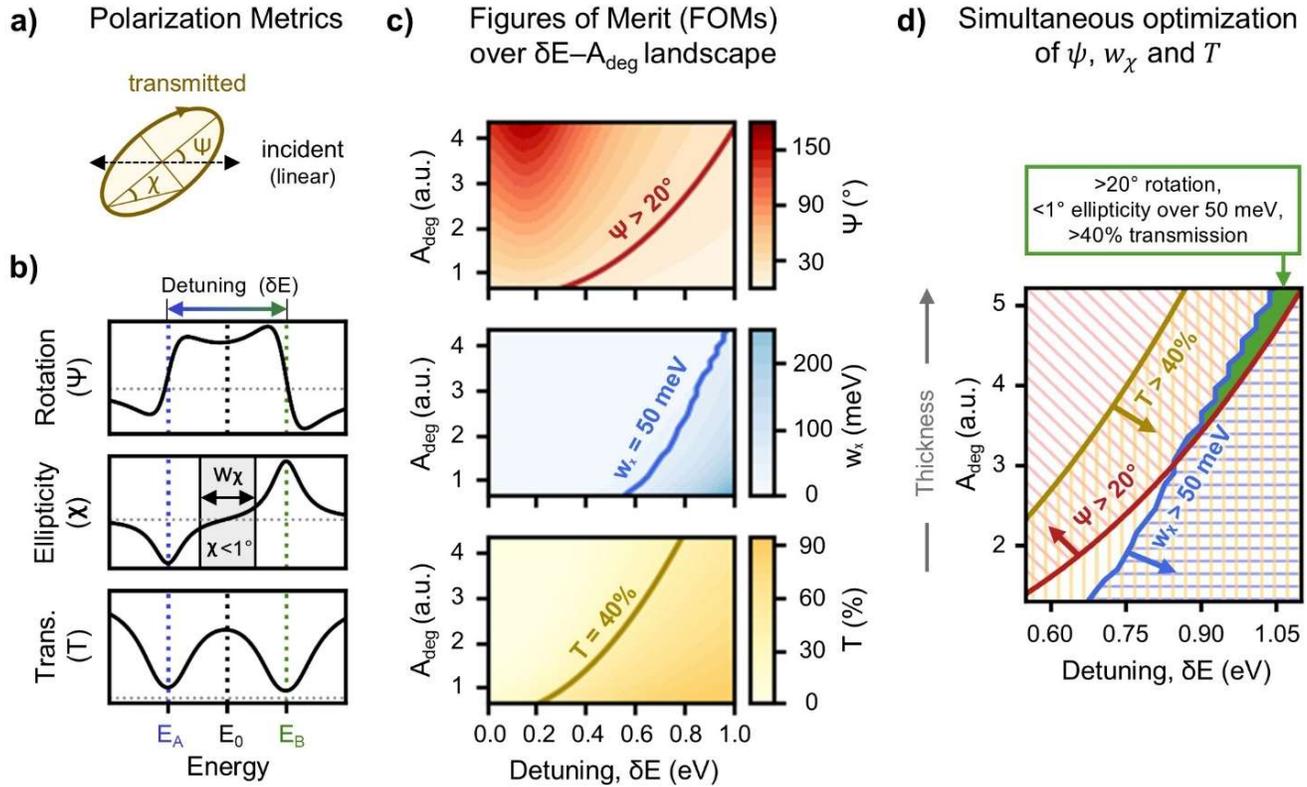

Figure 5: Quantifying optical rotation performance of the theorized ABAB non-degenerate heterostructure. a) Defining the polarization parameters of the transmitted radiation: Optical rotation ($\psi$) and ellipticity ($\chi$). b) Representative response to linearly polarized light of an ABAB structure where the chromophores A and B are detuned by δE. The rotation, ellipticity, and transmission inherit their salient features from the CB, CD, and absorption respectively. We use the following figures of merit (FOMs): Optical rotation ($\psi$) at $E_0$, the energy window where ellipticity is lower than 1°($w_\chi$, gray region) and the transmission at $E_0$; where $E_0$ is the midpoint between the two resonances. c) FOM colormaps plotted on a chromophore detuning - sample absorbance parameter space for a structure with $\theta = 60°$ and chromophore resonance FWHM $\gamma = 0.25 \, eV$. Here, the absorbance of the non-degenerate structure ($A_{deg}$) is the variable on the y-axis, which is directly proportional to the sample thickness. d) Overlaid contours from the three colormaps in c). The region with red hatching, above the red curve, denotes $\psi > 20°$; the blue region $w_\chi > 50 \, meV \, (12 \, THz)$ and the yellow region has a transmittance of greater than 40%. At higher thickness and chromophore detuning, these three regions overlap (represented with green), demonstrating that the proposed theoretical structure can simultaneously achieve large optical rotation (20°) with high transmission (40%) and low ellipticity (1°), over a window of 50 meV (12 THz).